\documentclass{epl}

\usepackage{amssymb}
\usepackage{amsmath}

\title{Non-equilibrium Magnetization Dynamics in the Fe$_8$ Single-Molecule Magnet Induced by High-Intensity Microwave Radiation}
\author{M. Bal\inst{1} \and Jonathan R. Friedman\inst{1} \and Y. Suzuki\inst{1,2} \and E. M. Rumberger\inst{3} \and D. N. Hendrickson\inst{3} \and N. Avraham\inst{4} \and Y. Myasoedov\inst{4} \and H. Shtrikman\inst{4} \and E. Zeldov\inst{4}}
\institute{
  \inst{1} Department of Physics, Amherst College, Amherst, Massachusetts 01002-5000\\
  \inst{2} Physics Department, City College of the City University of New York, New York, New York 10031\\
  \inst{3} Department of Chemistry and Biochemistry, University of California at San Diego, La Jolla, California 92093\\
  \inst{4} Department of Condensed Matter Physics, The Weizmann Institute of Science, Rehovot 76100, Israel
}
\pacs{75.50.Xx}{Molecular magnets}
\pacs{76.30.-v}{Electron paramagnetic resonance and relaxation}
\pacs{75.45.+j}{Macroscopic quantum phenomena in magnetic systems}

\begin{document}

\maketitle

\begin{abstract}
Resonant microwave radiation applied to a single crystal of the molecular magnet Fe$_8$ induces dramatic changes in the sample's magnetization. Transitions between excited states are found even though at the nominal system temperature these levels have negligible population.  We find evidence that the sample heats significantly when the resonance condition is met.  In addition, heating is observed after a short pulse of intense radiation has been turned off, indicating that the spin system is out of equilibrium with the lattice.
\end{abstract}

\section{Introduction}
Unlike classical magnets, single-molecule magnets, with spin values on the order of 10, have a discrete energy-level structure.  This allows for the observation of resonant tunneling between ``up" and ``down" orientations when energy levels are brought into resonance by an external magnetic field~\cite{33, 497, 81, 91}.  The discrete level structure also permits radiation to drive transitions between states.  Recent experiments show that microwave radiation can be used to enhance the rate for magnetization reversal~\cite{335, 283} and change the equilibrium magnetization~\cite{343,339,bal,petukhov,cage}.   Much of this research is aimed at studying the spin dynamics of single-molecule magnets in the presence of radiation with the goal of observing coherent dynamics such as Rabi oscillations and measuring the characteristic spin relaxation time $T_1$ and decoherence time $T_2$.  Measurements of these parameters are essential for determining whether single-molecule magnets are viable as potential qubits~\cite{ll}.

In this letter, we show that high-amplitude radiation can induce dramatic changes in the equilibrium magnetization of the molecular magnet Fe$_8$ when the radiation frequency matches the energy difference between various energy levels in the system.  This allows us to perform what is essentially a variant of electron-spin resonance spectroscopy on this system.  In addition, we find evidence that the radiation produces significant sample heating when the resonance condition is fulfilled. Using short radiation pulses, we observe spin heating effects even after the radiation has been turned off, an effect that indicates that the spin system and the lattice have been driven out of equilibrium.  These results imply that experiments seeking to study coherent spin dynamics in a radiation field need to be done very fast in order to circumvent heating effects.  

\begin{figure}[htb]
\centering
\onefigure[width=80mm]{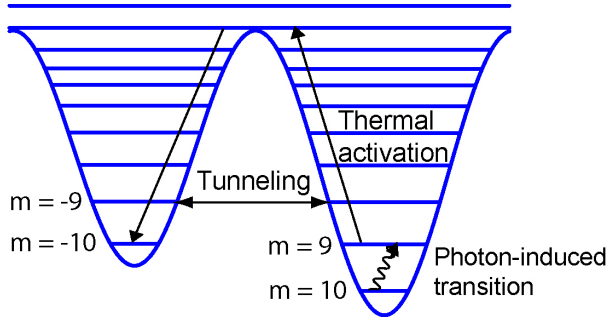} %\vskip 20 pt
\caption{Double-well potential and energy levels for the Fe$_8$ magnet.  The potential is drawn as a function of the angle $\theta$ between the spin and the easy (z) axis.  The left well corresponds to the spin pointing ``down" and the right corresponds to it pointing ``up".  The different processes involved in spin relaxation are illustrated schematically by the arrows.  Resonant microwave radiation can drive spins between neighboring states in a well (wavy arrow).  Phonons are responsible for thermal activation over the barrier.  Resonant tunneling can occur when the field tunes levels in opposite wells to align (horizontal arrow).  For clarity, only one photon resonance and one tunneling resonances are shown.
\label{doublewell} }
\end{figure} 

We study the single-molecule magnet Fe$_8$O$_2$(OH)$_{12}$(tacn)$_6$ (hereafter called Fe$_8$), which at low temperatures behaves as a single spin-10 system with the effective Hamiltonian

\begin{equation}
{\cal H} =  - DS_z^2  + E\left( {S_x^2  - S_y^2 } \right) + C\left( {S_ + ^4  + S_ - ^4 } \right) - g\mu _B \vec S \cdot \vec H
\label{Ham}
\end{equation}
where the anisotropy constants $D$, $E$, and $C$  are 0.292 K, 0.046 K, and -2.9 x 10$^{-5}$ K, respectively, and $g$ = 2~\cite{162, 188, 184, mukhin}. The first term in the Hamiltonian makes the z axis the preferred axis of orientation, the so-called easy axis.  The second and third terms break the rotational symmetry of the Hamiltonian and result in tunneling between the otherwise unperturbed states.  The spin's energy can be described using a double-well potential, shown in Fig.~\ref{doublewell}, where the bottom of the left (right) well corresponds to the spin pointing along (antiparallel to) the easy axis.  A $\sim$25 K barrier~\cite{91} separates the ``up" and ``down'' orientations and the $2S+1=21$ energy levels correspond to different orientations of the magnetic moment.  A magnetic field applied along the easy axis tilts the potential.  

In the absence of radiation, Fe8 (similar to other systems of single-molecule magnets) relaxes by a combination of thermal activation and tunneling above $\sim$0.4 K and by pure quantum tunneling below this temperature~\cite{91}.  In the higher-temperature regime, the relaxation rate $\Gamma$ has an Arrhenius-law dependence on temperature:  $\Gamma = \omega_0 e^{-U_{eff}/k_B T}$, where $\omega_0 \sim 10^{7}$ rad/s, and $U_{eff}$ is the effective barrier height, which can be modulated by tunneling.

\section{Experimental Details}
In previous work~\cite{bal}, we used a low-power radiation source and found small changes (less than 1\%) in the magnetization whenever the radiation frequency matched the energy difference between the ground and first excited states.  Here we increased the radiation amplitude both by using a high-power backward-wave oscillator source and a  rectangular cavity with a Q as high as 2500.  The largest cavity radiation field is $\sim$0.5 G, sufficient to create substantial radiative transition rates between energy levels~\cite{ll}.  We mounted an Fe$_8$ single crystal inside the cavity.  One end of the cavity had a small opening to allow the coupling of radiation from the attached waveguide.  A Hall-bar detector was mounted outside the cavity next to the sample position.  We mounted the sample with its $a$ axis parallel to the applied field direction, which results in the easy axis being tilted $\sim$16$^\circ$ from the field direction~\cite{332}. 

\section{Results and Discussion}
\begin{figure}[htb]
\centering
\twofigures[width=70mm]{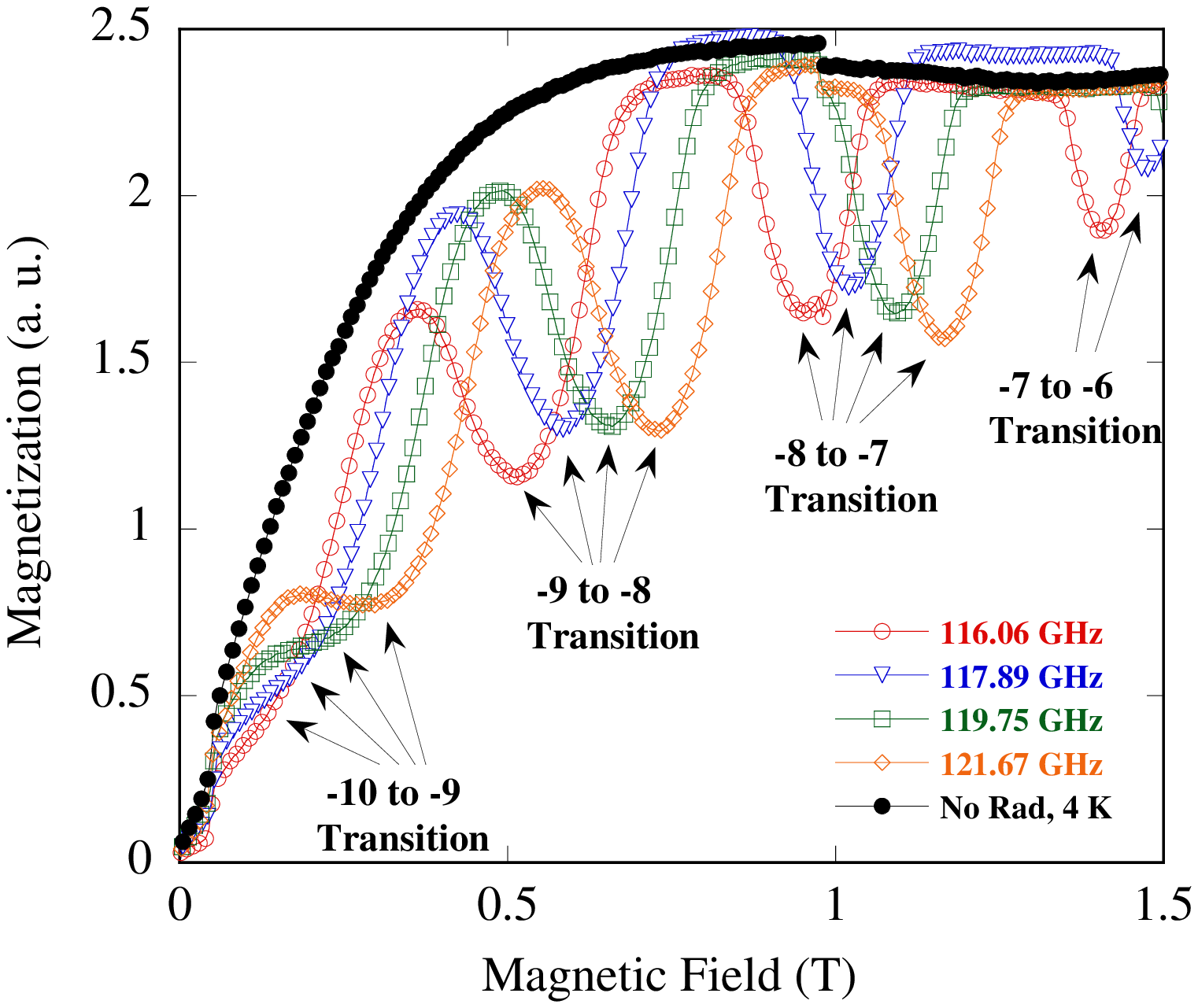}{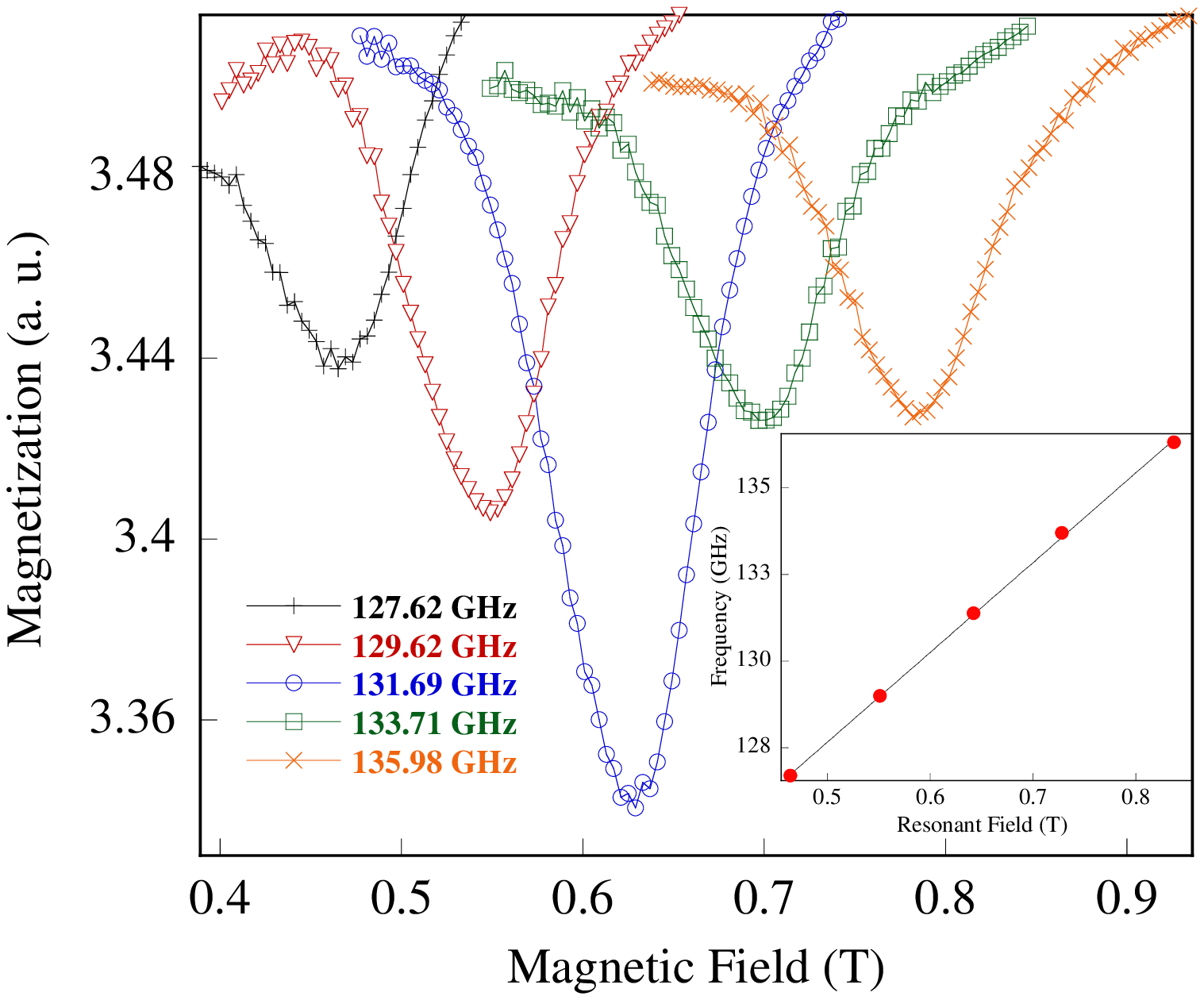} %\vskip 20 pt
\caption{Equilibrium magnetization as a function of field with and without the presence of radiation, as indicated.  The dips in magnetization occur when the applied radiation frequency matches the transitions between states with the indicated magnetic quantum numbers.  The applied radiation power and cavity Q were 12--17 mW and 1500--2500, respectively, depending on the frequency.  The abrupt jumps in the data, most conspicuous just below 1 T, are due to instrumental artifacts.}
\label{peaks} 
\caption{Magnetization dips for the 10-to-9 transition at several radiation frequencies, as indicated.  The radiation power was between 2.5 and 3.7 mW, depending on frequency, and the cavity Q value was several hundred.  The inset shows the transition frequency versus the field positions of the dips.  A straight line fit allowed us to extract a value for the anisotropy parameter D of 0.2944 $\pm$ 0.0005 K.}
\label{zeeman} 
\end{figure}

Fig.~\ref{peaks} shows the measured magnetization of the sample both in the absence of radiation and when high-power (12--17 mW) radiation of various frequencies has been applied.  The Q of the cavity is between 1500 and 2500, depending on the mode excited.  At each frequency, the radiation induces a large dip in the sample's magnetization at certain values of magnetic field.  At these fields, the frequency of the radiation matches the energy difference between two neighboring levels in, say, the right well in Fig.~\ref{doublewell}, resulting in the absorption of radiation and a corresponding change in magnetization.  The observed magnetization changes are so large that they cannot be attributed to simply moving population between the two states involved in the resonance.  Instead, much of the population must be transferred to the opposite well, through thermal and/or quantum processes.

Our results are consistent with numerous spectroscopic studies~\cite{188, 184, mukhin, park} on this material.  To confirm the accuracy of our measurement technique as a spectroscopic tool, careful data was taken at somewhat lower power (and with a lower cavity Q of a few hundred) of the 10-to-9 transition at several different frequencies, as shown in Fig.~\ref{zeeman}.  The inset shows the applied radiation frequency  as a function of the measured field positions of the peaks.  The straight-line fit confirms the expected Zeeman dependence, the last term in Eq.~\ref{Ham}.  The intercept of the fit yields a value for the anisotropy parameter D of 0.2944 $\pm$ 0.0005 K, in good agreement with the accepted values for this material.

One surprising feature of the data in Fig.~\ref{peaks} is the observation of transitions between excited states.  The nominal temperature of our cryostat is $\sim$2 K during these experiments.  When radiation is not resonantly absorbed by the sample, it nevertheless heats the cavity (which in turn warms the sample) to $\sim$4 K, as evidenced by the fact that the data in Fig.~\ref{peaks} taken at 4 K in the absence of radiation approximately coincides with the data taken in the presence of radiation when the field is off resonance.  Even at this elevated temperature, transitions between excited states should be weak, if observable at all, since the thermal populations of the excited states are small.  We believe that a positive-feedback mechanism is at play that creates excess sample heating when the radiation is resonant with transitions between levels.  To wit, even with a small population in an excited state the sample will absorb radiation when on resonance and some of this population will be put into a higher-energy state.  In decaying from this state, a spin will release a phonon rather than a photon due to the much higher phonon density of states.  These phonons rapidly thermalize, raising the temperature of the sample.  This increases the population of the excited states, and, in turn, the number of photons absorbed, fuelling the feedback mechanism and driving the system to a much higher temperature than its environment.

\begin{figure}[htb]
\centering
\onefigure[width=80mm]{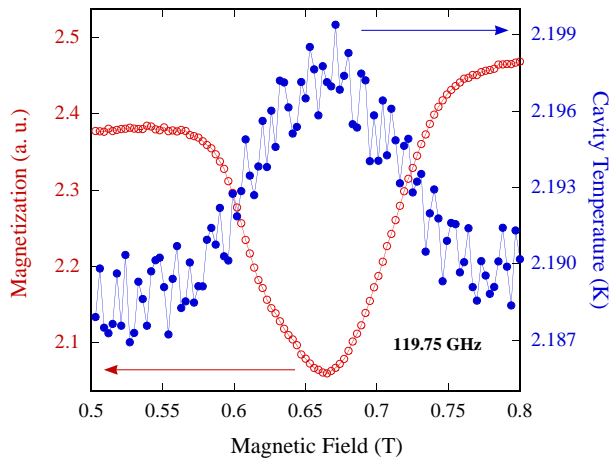} %\vskip 20 pt
\caption{Simultaneously measured sample magnetization and cavity temperature as the magnetic field is swept through resonance for the 9-to-8 transition for the applied 119.75 GHz radiation.  The applied radiation power was $\sim$3.2 mW and the cavity Q was $\sim$2400.  The cavity temperature was measured with a thermometer mounted outside the cavity, while the sample was inside.  The observed $\sim$10 mK temperature rise reflects a much larger change in the sample temperature.
\label{temp} }
\end{figure}

To confirm that the sample heats additionally when on resonance, a thermometer was attached to the outside of our cavity about a centimeter away from the sample position and  the temperature was monitored as  the field was swept through one of the resonances (9 to 8).  The thermometer registered a clear temperature rise of about $\sim$10 mK when the field tuned through the resonance, as shown in Fig.~\ref{temp}.  The specific heats of the cavity and sample are within an order of magnitude of each other at the experimental temperatures~\cite{AgCp,Fe8Cm}.  However, the cavity mass ($\sim$10g) is many orders of magnitude larger than the sample's ($\sim$0.5$\mu$g).  Therefore, the observed cavity temperature rise probably corresponds to a several-Kelvin increase in the sample's temperature.  

\begin{figure}[htb]
\centering
\onefigure[width=135mm]{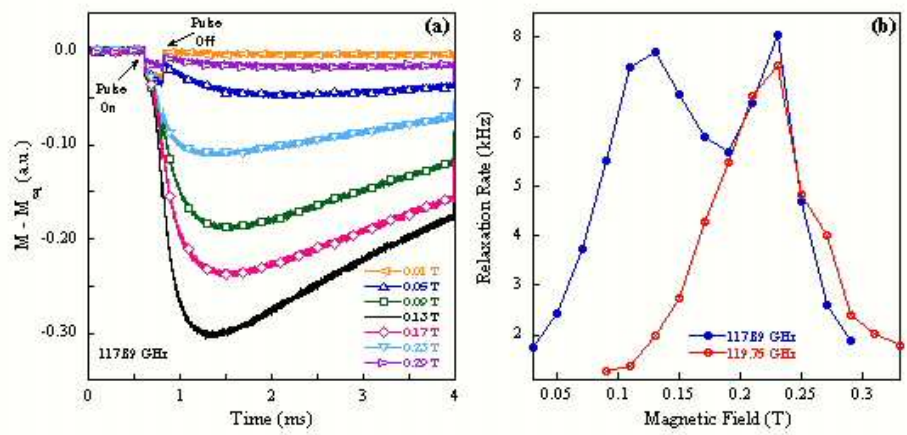} %\vskip 20 pt
\caption{(a) Magnetization change as a function of time during and after a 0.2-ms pulse of 117.89 GHz radiation at several values of magnetic field, as indicated, and a cryostat temperature of 1.8 K.  The decrease in magnetization after the radiation has been turned off indicates that the spin system is out of equilibrium with the lattice.  The abrupt jumps in the data at the beginning and end of the radiation pulse are instrumental artifacts.  (b)  Relaxation rate as a function of magnetic field for the data in (a) and similar data taken at 119.75 GHz.  The relaxation rate was determined by fitting the decay after the radiation pulse to an exponential.
\label{pulse} }
\end{figure}

To better understand how this heating takes place, we performed a time-resolved study of the magnetization by applying a single 0.2-ms pulse~\cite{note1} of 117.89 GHz radiation into a cavity with a low Q ($\sim$100 --200). The change in magnetization from its equilibrium value is plotted as a function of time at several fixed magnetic fields in Fig.~\ref{pulse}a. The abrupt jumps that occur when the radiation is turned on or off are instrumental artifacts. Ignoring these, the magnetization of the sample decreases during the radiation pulse due to photon-induced transitions from the ground state to the first excited state as well as heating. However, after the radiation pulse is turned off, the magnetization continues to decrease for an additional fraction of a millisecond and reaches a minimum~\cite{note}. The decrease is largest at 0.13 T, the field at which the 10-to-9 transition occurs in Fig.~\ref{peaks} for this frequency.  Since energy is no longer being supplied to the system, one might expect the magnetization to increase as the system cools.  In contrast, it appears that the spins absorb radiation during the radiation pulse, releasing phonons, which heat the lattice as they thermalize.  The magnetization then continues to decrease as the spin system comes to equilibrium with the lattice in the (much longer) characteristic time required for the spins to cross the barrier (Fig.~\ref{doublewell}). Thus, although photons are being absorbed by the spins, thermalization occurs first in the lattice and then in the spin system.  

We checked this hypothesis by fitting the the region between the end of the pulse and the magnetization minimum to an exponential.  The relaxation rate (Fig.~\ref{pulse}b) determined this way shows two peaks:  one at a field of $\sim$0.13 T, the 10-to-9 transition resonance where the amount of heating is greatest, and another at $\sim$0.23 T, at which the first tunneling resonance beyond zero field occurs for this system (illustrated in Fig.~1).  Fig.~\ref{pulse}b also shows the results of a similar analysis of data taken at 119.75 GHz, where the field at which the 10-to-9 radiation transition nearly coincides with that for resonant tunneling, causing the two peaks to merge into one.  These data indicate that the magnetization decrease after the radiation pulse appears to be controlled by the relaxation rate associated with the spin system, which is increased either by radiatively raising the temperature or by tunneling.  This supports the notion that the decrease in magnetization after the pulse results from the the spin system returning to equilibrium with the lattice, which has been indirectly heated by the radiation. These results may have some bearing on the slow recovery times measured by del Barco {\em et al.}~when the Ni$_4$ molecular magnet is irradiated by microwaves at low temperatures~\cite{339}.

\begin{figure}[htb]
\centering
\onefigure[width=80mm]{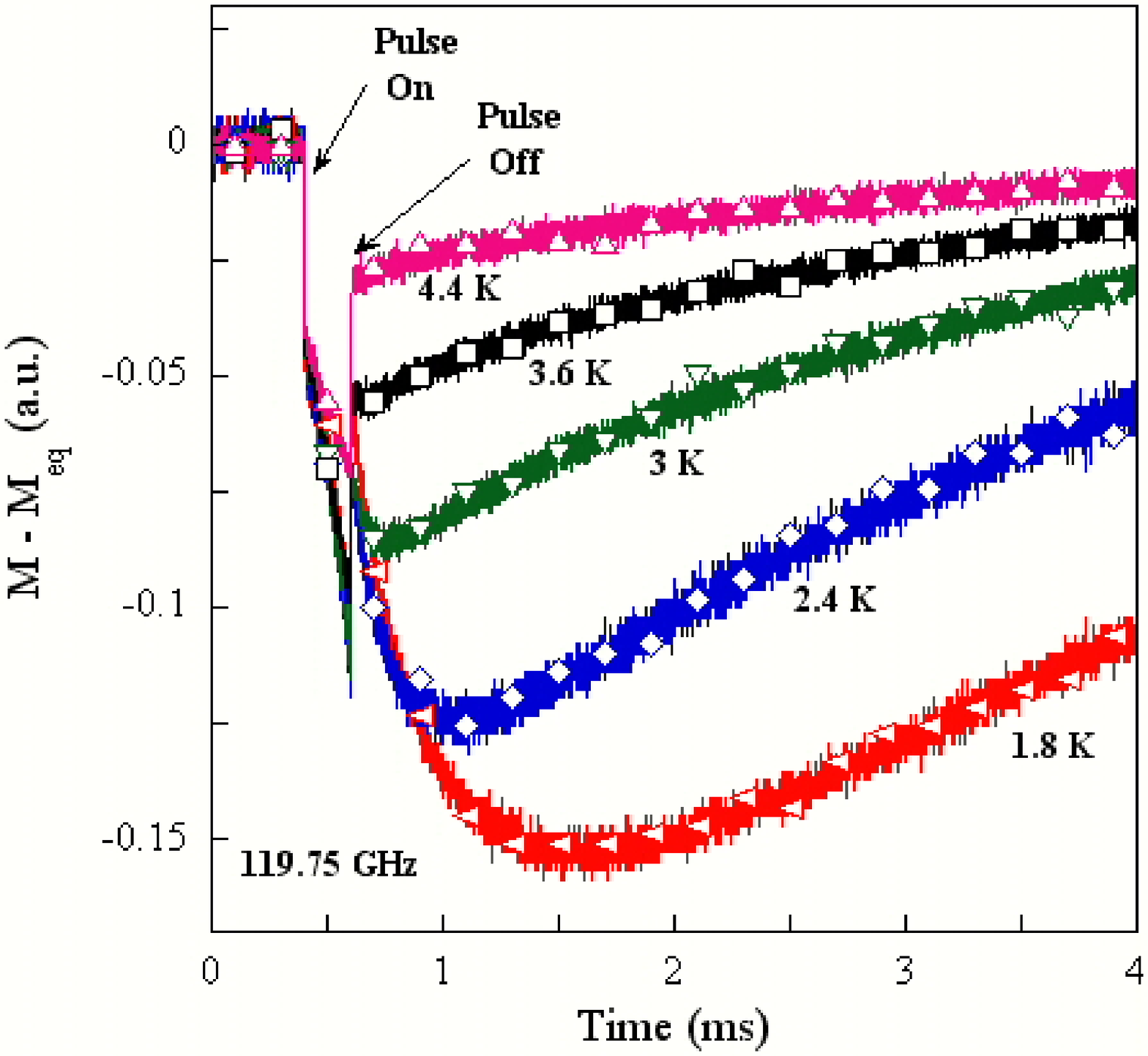} %\vskip 20 pt
\caption{Magnetization change induced by radiation pulses as a function of time  at several different cryostat temperatures when the magnetic field is fixed at 0.17 T.  The induced magnetization change after the pulse becomes smaller and the response faster as the temperature is raised, consistent with the faster relaxation rate of the spin system at higher temperatures.
\label{pulsetemp} }
\end{figure} 

The change in magnetization due to a radiation pulse at different temperatures is plotted in Fig.~\ref{pulsetemp}. The decrease in magnetization after the pulse diminishes for increasing temperature up to 3.6 K, above which there is no longer a decrease in magnetization immediately after the pulse. This is consistent with our picture since at higher temperatures the spins relax faster (in accordance with an Arrhenius law) and thereby equilibriate with the lattice more quickly. In addition, the  temperature increase of the lattice  during the radiation pulse is expected to be smaller at higher temperatures  because of the lattice's larger specific heat, reducing the magnitude of the observed effect.   

\section{Conclusions}
We find that resonant high-power microwave radiation can induce significant changes in the equilibrium magnetization of the molecular magnet Fe$_8$ when the radiation is on resonance with transitions between states with different magnetic quantum numbers.  We observe transitions between excited states even when the thermal populations of these states is very small, indicating a substantial amount of sample heating when on resonance, as we have confirmed by monitoring the cavity temperature. When pulses of radiation are applied, the spin system and the lattice are driven out of thermal equilibrium, with recovery determined by the spin relaxation time. Any experiment probing photon-induced magnetization dynamics between only two levels of the spin system needs to be done at much shorter time scales.

\acknowledgments
We thank M. P. Sarachik, K. M. Mertes, J. Tu, E. M. Chudnovsky and J. Tejada for useful conversations.  We are indebted to D. Krause, P. Grant and R. Chaudhuri for their technical contributions.  Support for this work was provided by the US National Science Foundation under grant number CCF- 0218469, the Alfred P. Sloan Foundation, the Center of Excellence of the Israel Science Foundation, and the Amherst College Dean of Faculty.


\begin{thebibliography}{0}
\bibitem{note1}  We also did experiments with pulses ranging in duration from 1 $\mu$s to 2 ms and obtained qualitatively similar results.  When pulse durations were short, the induced magnetization change was barely measurable.  In these experiments we were restricted to time scales $\gtrsim$1 $\mu$s, the response time of our Hall detector.  Experiments at shorter time scales, at which coherent radiation-induced dynamics may be observable~\cite{ll}, are in progress.
\end{thebibliography}
\end{document}